# A Semi-automated Peer-review System

Bradly Alicea
bradly.alicea@ieee.org
Orthogonal Research

**Abstract**

A semi-supervised model of peer review is introduced that is intended to overcome the bias and incompleteness of traditional peer review. Traditional approaches are reliant on human biases, while consensus decision-making is constrained by sparse information. Here, the architecture for one potential improvement (a semi-supervised, human-assisted classifier) to the traditional approach will be introduced and evaluated. To evaluate the potential advantages of such a system, hypothetical receiver operating characteristic (ROC) curves for both approaches will be assessed. This will provide more specific indications of how automation would be beneficial in the manuscript evaluation process. In conclusion, the implications for such a system on measurements of scientific impact and improving the quality of open submission repositories will be discussed.

## Introduction

What makes for a good academic paper? The answer is often more subjective than we like to think. One common answer involves a peer-reviewed consensus. But peer-review can often be subjective, and yields results that do not lend easily to consensus [1]. Perhaps a high degree of selectivity can provide a benchmark for quality. Yet even highly-selective publication venues can suffer from retractions and uneven quality. Part of the problem with the standard model of peer review (a small number of blinded reviews coordinated by an editor) is that the entire process represents a sparse sampling of the subject space. To compensate for this, manuscripts with multiple novel features [2] will often be viewed in an over-critical manner, perhaps even being discarded as fraudulent.

How can the vagaries of opinion, aversion to novelty, and bias for selectivity be overcome? This is an important issue with the rise of open-access publication [3]. One way is to make rear-guard improvements to the existing peer-review system. Another option is to build an automated system that filters out fraudulent papers and papers of low quality. Superficially, this system would function like a human-assisted fraud detection system [4], and would provide a uniform set of decisions given a set of assumptions.

## Model Assumptions

There are several assumptions that go into the semi-automation model. The first assumption involves a **natural fraud rate**. This is the frequency at which intentionally fraudulent papers are encountered. The second assumption involves an **acceptance rate** (or rejection rate) that may overlap with the natural fraud rate. The third assumption is that the true number of non-fraudulent manuscripts can be better estimated by establishing a **degree of novelty**. These assumptions can be combined to come up with false positive and false negative rates for specific datasets (see Table 1).

**Table 1.** Example of false positives (FPs) and false negatives (FNs) given a certain natural fraud rate (10%) and acceptance rate (20%).

|  | **Accept (20%)** | **Reject (80%)** |
|---|---|---|
| **Legitimate (90%)** | TP (15%) | FN (75%) |
| **Fraud (10%)** | FP (5%) | TN (5%) |

### Design Features

The design of the automated peer review system utilizes an adaptive filtering architecture (see Figure 1). This consists of a classifier algorithm, a discriminability measure, and ROC curves to characterize the performance of both existing and automated systems with respect to how strategies can eliminate fraud or (in some cases) hamper innovation in the literature. Thus, while the goal is not to build a fraud-proof system for evaluating manuscripts, an automated peer review system can help us understand how to eliminate human biases and more efficiently evaluate submissions to the scientific literature.

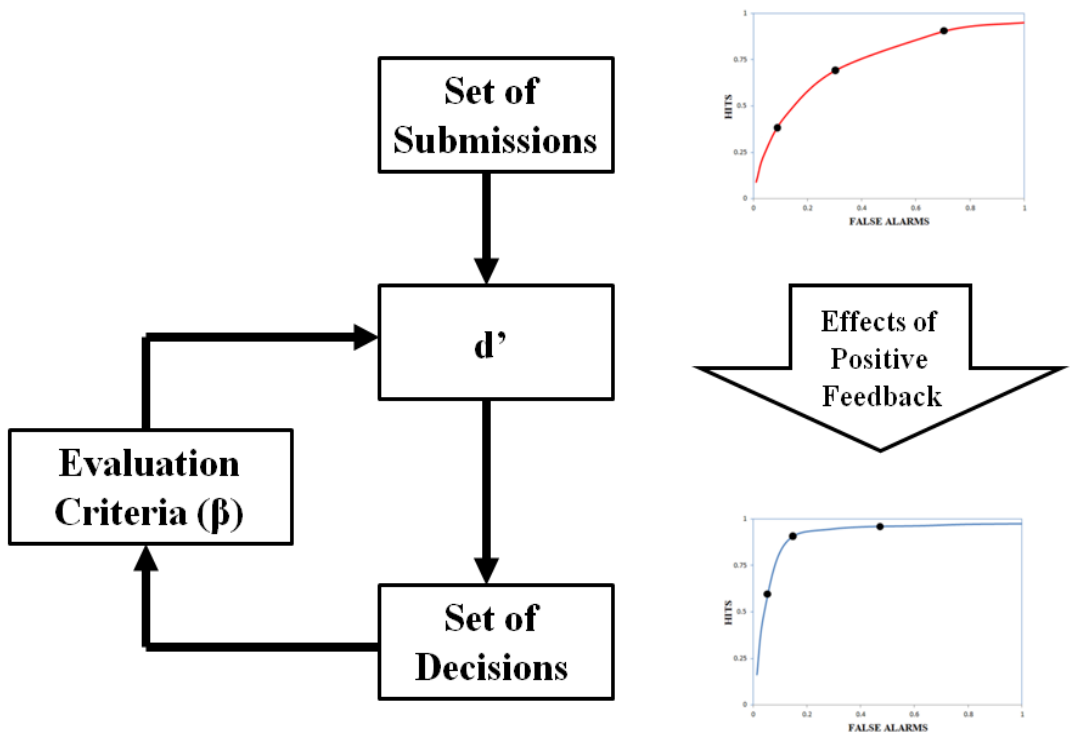

**Figure 1.** Process of semi-automated decision-making that utilizes an adaptive filter.

### System Design Features

There are three design features that are included in this automated system. The first involves excluding all fraudulent work by discovering the natural rate of fraudulent submissions. This includes plagiarized works, and works which include errors in analysis or basic logic. For example, papers might be flagged for graphical and analytical (e.g. statistical artifacts) errors.

The second is to ensure the inclusion of novelty, particularly in cases where it may be confused for low quality or fraudulent work. The third feature is to approach the paper being reviewed from multiple points-of-view. This consists of rule systems which are intended to transcend specific disciplinary (or experiential) biases, but also overlap in a way that allows a consensus to be reached.

In Figure 1, the *d'* (discriminability) and *β* (bias) parameters serve as a filter and adaptive calibration mechanism, respectively. The d' parameter is the ability of a classifier (set of minimal quality and anti-fraud criteria) to minimize false positive and false negatives with respect to random expectation. The *β* parameter is a shift in the decision-making criteria, which results in either a more permissive or selective strategy. A given set of decision and adjustment of the evaluation criterion can be used to adjust the discriminability value during the next round of evaluations.

**ROC Curves**
To characterize the effects of selectivity and evaluation criteria (the *d'* parameter and associated feedback loop), we can use ROC curves. Figure 2 demonstrates the outcomes for both a traditional peer review and semi-automated review system. The ROC curve characterizes general tendencies for different strategies, in this case the relative degree of selectivity for a prestigious journal, a field-specific journal, and an open-submission repository.

Figure 2 provides a schematic view of performance for traditional vs. semi-automated peer review using the same criteria. In terms of discriminability, the semi-automated approach is expected to outperform traditional peer-review. This should especially be true of papers with a high degree of novelty. In turn, this should allow for a higher acceptance rate at a lower cost in terms of quality.

**Set of Minimal Criteria**
The parameter value for d' is determined by the function of a classifier. The classifier is a set of minimal criteria that are used to classify a given manuscript as fraudulent, below-threshold quality, or acceptable but needing work. This can be done in two ways. The first is by using a rule-based sort, which sorts papers by their location in an *n*-level contingency tree. The other way is to generate an *n*-dimensional score using the same criteria, and then apply a discriminant classifier to the manuscript scores. All manuscripts classified "accept" and "reject" are then double-checked by human experts using the same criterion. This human expert supervision step provides us with the final classifications of "true positive", "false positive", "true negative", and "false negative".

## Discussion
One of the things an ROC framework reveals is the costs of rooting out fraud using naive (non-automated) techniques. One such naive strategy for rooting out fraud is to implement greater selectivity. While this has a heuristic benefit, it also serves as a *de facto* conservative bias against all manuscripts that do not meet a rigid criterion. This is because selectivity favors well-established people and techniques. The costs of homogenization process are both social and intellectual. By contrast, automation allows us to distinguish between novelty and a lack of quality (e.g. fraud). See the Methods section for more details.

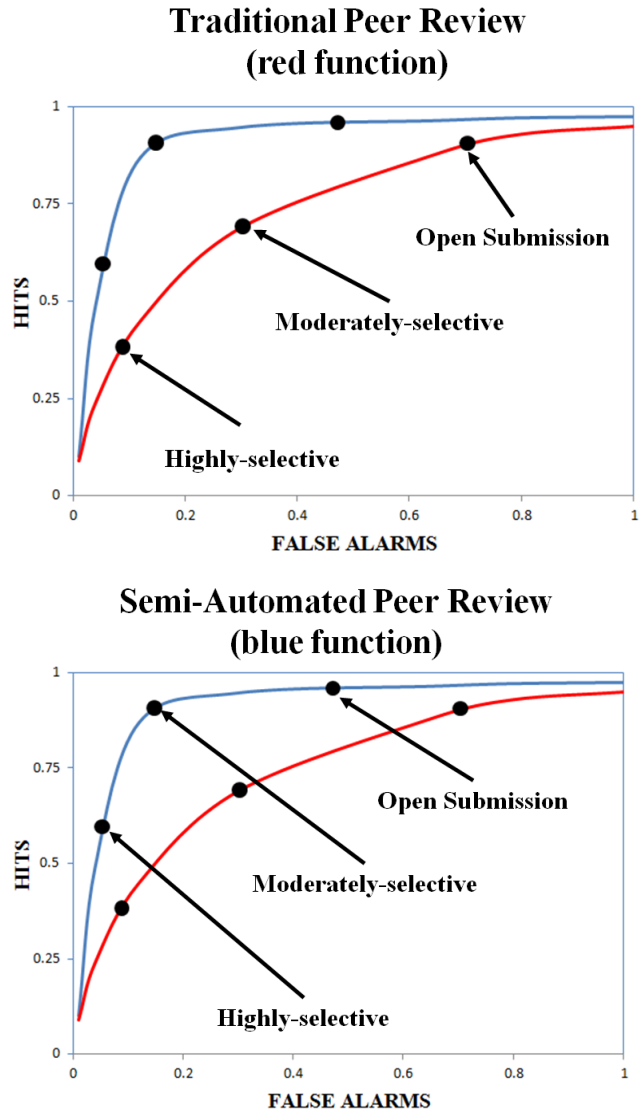

**Figure 2.** Example of ROC curves for three datasets. Curves are generated using pseudo-data, and represent general expected tendencies for traditional (red) and semi-automated (blue) manuscript evaluation.

One existing alternative to traditional peer-review is the Altmetric approach [5]. With Altmetrics, articles can be assessed using the power of crowdsourcing. However, this still necessitates both good faith and the ability to detect fraud amongst the people evaluating the manuscript. Altmetrics also introduces the concept of an article's potential for influence (e.g. relative impact) [6], which is opportunity for future development of the proposed automated peer-review system. Indeed, an Altmetric approach would be quite suitable as the human training component of the proposed system.

Even in the case of open submission repositories, there are criteria for suitability. The arXiv repository [7] uses a volunteer-run committee of subject experts to manually evaluate manuscripts. The criteria for minimal selectivity and bias involve a general assessment of an article's scientific interest, relevance, and value. Fraud detection and quality control includes the identification of duplicate work, an excessive personal publication rate, and copyright violations. Open submission is another area in which an automated system could help keep repositories free of low quality and/or fraudulent papers. Yet for all applications, this semi-automated system provides a opportunity to review manuscripts in a transparent, informed, and unbiased manner.

## Methods

**Discriminability measure**

Discriminability is measured by modeling the signal (classifier performance) and noise (random expectation) as overlapping Gaussian probability distributions. Discriminability (d') can be calculated by calculating the overlap in the probability distributions

$$d' = \int f(S) - \int f(N) \quad [1]$$

where f(S) is the distribution that describes the signal, and f(N) is the distribution that describes noise. Discriminability determines the overall performance of the specific evaluation. In an adaptive context, d' serves to select the overall acceptance rate. This measurement can also be used to compare between conventional peer-review systems.

**ROC curves**

ROC curves characterize all the possible true positive (hits) and false positive values for a specific d' value. Bias ($\beta$) is characterized by [8] as shifts in the acceptance criterion

$$\beta = \frac{\frac{1}{\sqrt{2\pi}} \exp\{\frac{1}{-2} D^2\}}{\frac{1}{\sqrt{2\pi}} \exp\{\frac{1}{-2} B^2\}} \quad [2]$$

$$\beta = \exp\{d' \times C\} \quad [3]$$

**C > 0 = conservative and C < 0 = permissive**

where B is the threshold criteria (e.g. *a priori* acceptance rate) for an acceptance, and C characterizes changes in the strategy (e.g. highly-selective vs. open-submission). $\beta$ can also serve as a likelihood ratio between parameters D and B [8]. Over time, the value of $\beta$ will converge upon one that results in maximizing the number of acceptance and minimizing the number of false positives (unacceptable manuscripts).

**Degree of Novelty**

The degree of novelty is the information content of the rulesets states (for the *n*-level contingency tree) or scores (for the discriminant classifier) that covers the methods and conventional approaches part of the rulesets

$$N = \sum_{j=n} p_i \log p_i \quad [4]$$

where $p_i$ is the probability of a given score value or state (*i*) for each rule ($j_1, j_2,..j_n$). To make the final decision, the degree of novelty will be used to weight the final classifications for each manuscript. Cases of moderate to high degrees of novelty will serve as a flag for the human supervisor to further investigate the paper's merits.

**Classifiers**

The rulesets consist of rules that cover the following components of a manuscript: methods, reasoning, plagiarism check, references and self-citation, conventionality of approach, and graphical consistency/analytical artifacts. The rulesets can either result in a binary string or a composite score, and can include as many rules in each category as deemed use-appropriate. All code and pseudo-code are located at Github (http://www.github.com/balicea/semi-auto-peer-review).